\documentclass[fleqn,a4paper]{article}

\begin{document}

\title{Neville's algorithm revisited}
\author{M.\ de Jong\\
  {\small NWO-I, Nikhef, PO Box 41882, Amsterdam, 1098 DB Netherlands} \\
  {\small Leiden University, Leiden Institute of Physics, PO Box 9504, Leiden, 2300 RA Netherlands}
}
\maketitle

\begin{abstract}
  Neville's algorithm is known to provide an efficient and numerically stable solution for polynomial interpolations.
  In this paper, an extension of this algorithm is presented which includes the derivatives of the interpolating polynomial.
\end{abstract}

\section{Introduction}

In general, polynomial interpolation is based on the unique polynomial, $P_{N}(x)$, of $N$ degrees
which exactly goes through the $N+1$ points $(x,y)$ of some function $f(x)$.
Here, $x$ and $y$ refer to the abscissa and ordinate values of the function, respectively.
By using Neville's algorithm, this polynomial is evaluated at some arbitrary $x$ in the following way.
Let $P_{i,j}$ denote the polynomial of degree $N=j-i$ which goes through the points $(x_k,y_k)$ for $k = i, i + 1, \ldots, j$.
The $P_{i,j}$ should then satisfy the following recurrence relations:

\begin{eqnarray}
  \label{eq:neville-1}
  P_{i,i}(x) & = & \makebox[0.55\textwidth][l]{$\textstyle y_{i}$}    0 \le i \le N      \\
  \label{eq:neville-2}
  P_{i,j}(x) & = & \makebox[0.55\textwidth][l]{$\frac{\textstyle 
      (x_j-x) P_{i,j-1}(x) \;+\;
      (x-x_i) P_{i+1,j}(x)
    }{\textstyle x_j-x_i}$}                                          0 \le i < j \le N  
\end{eqnarray}

\noindent
These recurrence relations lead to the $P_{0,N}(x)$ which is the interpolated value of $y=f(x)$.
An implementation of these recurrence relations is for example presented in reference \cite{ref:numerical-recipes}.
It may be useful to also know the derivatives of the interpolating polynomial.
For instance, one could identify a (local) maximum or minimum based on the values of the first and second derivatives.
One could then also employ Newton-Raphson method to efficiently find the abscissa $x$ for which $f(x)$ has a certain value.

\section{Extension}

The recurrence relations that lead to the ordinate value $y=f(x)$ at a given $x$ can readily be extended to
include the derivatives of interpolating polynomial.
By application of the so-called chain rule for differentiating compositions of functions to
the equations \ref{eq:neville-1} and \ref{eq:neville-2}, one obtains:

\begin{eqnarray}
  \label{eq:neville-extended-1}
  P_{i,i}^{n}(x) & = & 0     \\
  \label{eq:neville-extended-2}
  P_{i,j}^{n}(x) & = & \frac{(x_j-x)P_{i,j-1}^{n+1}(x) \;-\; P_{i,j-1}^{n}(x) \;+\;  (x-x_i)P_{i+1,j}^{n+1}(x) \;+\; P_{i+1,j}^{n}(x)}{x_j-x_i}
\end{eqnarray}

\noindent
where $n$ refers to the $n^{th}$ derivative of the interpolating polynomial.
The same constraints to the indices $i$ and $j$ apply as in equations \ref{eq:neville-1} and \ref{eq:neville-2}.
These recurrence relations lead to the different $P_{0,N}^n(x)$
which are the interpolated values of $f^n(x)$, respectively.

\section{Tests}

As a test, the extension of Neville's algorithm is first applied to a polynomial function.
In this case, the derivatives of the interpolating function should 
--to a high precision-- be equal to those of the original function.
In the following, a polynomial function of the third degree is considered, namely:

\begin{eqnarray}
  \label{eq:polynomial}
  f(x) = a_0 + a_1x + a_2x^2 + a_3x^3 
\end{eqnarray}

\noindent
The coefficients are arbitrarily set to $a_0 = a_1 =  a_2 = a_3 = 1$ and 
the considered range of abscissa values is $\left[-1,+1\right]$.
The values of the function and its derivatives evaluated at $x=0$ are listed in table \ref{tab:test-1}.
For the interpolation, $11$ points $(x,y)$ have been evaluated at equidistant abscissa values of $x = \{ -1, -0.8, \ldots, +0.8, +1 \}$.
The values of the interpolating polynomial and its derivatives have also been evaluated at $x=0$
using the equations \ref{eq:neville-extended-1} and \ref{eq:neville-extended-2}.
The thus obtained values are also listed in table \ref{tab:test-1}.
As expected, the latter values are equal to the value of the original function.

\begin{table}[h!]
  \begin{center}
  \begin{tabular}{|l|c|c|c|c|}
    \hline
    \rule{0pt}{12pt} &   $f(x)$   &   $f^1(x)$   &   $f^2(x)$   &   $f^3(x)$   \\
    \hline
    original         &     1      &      1       &      2       &      6       \\
    \hline
    calculated       &     1      &      1       &      2       &      6       \\
    \hline
  \end{tabular}
  \end{center}
  \caption{
    Function values of the polynomial from equation \ref{eq:polynomial} evaluated at $x=0$ (row ``original'') and
    those calculated using the equations \ref{eq:neville-extended-1} and \ref{eq:neville-extended-2} (row ``calculated'').
    The superscript at $f^n(x)$ refers to the $n^{th}$ derivative of $f(x)$.
    \label{tab:test-1}
    }
\end{table}

For the second test, 1,000,000 random abscissa values, $x$, have been used which were uniformly generated between $\left[-1,+1\right]$.
For each $x$, the function values of the original polynomial and those of the interpolating polynomial are compared.
The average, the RMS and the maximum of the differences between the function values are listed in table \ref{tab:test-2}.

\begin{table}[h!]
  \begin{center}
  \begin{tabular}{|l|r@{}l|r@{}l|r@{}l|}
    \hline
    \rule{0pt}{12pt} & \multicolumn{2}{c|}{average} & \multicolumn{2}{c|}{RMS} & \multicolumn{2}{c|}{maximum} \\
    \hline
    $f(x)$           &          -2.&1e-17           &          1.&8e-16        &           1.&3e-15          \\
    $f^1(x)$         &          -2.&6e-17           &          8.&5e-16        &           7.&1e-15          \\
    $f^2(x)$         &           2.&3e-15           &          8.&7e-15        &           6.&7e-14          \\
    $f^3(x)$         &           1.&9e-14           &          6.&3e-14        &           5.&9e-13          \\
    \hline
  \end{tabular}
  \end{center}
  \caption{
    The average, RMS and maximum of the differences between
    the function values of the original polynomial and those of the interpolating polynomial.
    The superscript at $f^n(x)$ refers to the $n^{th}$ derivative of $f(x)$.
    \label{tab:test-2}
    }
\end{table}

\noindent
As can be seen from table \ref{tab:test-2}, the results are accurate to a numerical precision of better than $10^{-12}$.

A further test is done using the function $\sin(x)$.
For the interpolation, $21$ points $(x,y)$ have been evaluated at equidistant abscissa values between $0$ and $2\pi$.
For this test, 1,000,000 random abscissa values, $x$, have been used which were uniformly generated between $\left[0,2\pi\right]$.
The RMSs of the differences between the function values of $\sin(x)$ and
those of the interpolating polynomial are listed in table \ref{tab:test-3}
for different degrees of the interpolating polynomial.

\begin{table}[h!]
  \begin{center}
  \begin{tabular}{|l|r@{}l|r@{}l|r@{}l|r@{}l|}
    \hline
    degree:          & \multicolumn{2}{c|}{2}   & \multicolumn{2}{c|}{3}   & \multicolumn{2}{c|}{4}   & \multicolumn{2}{c|}{5}   \\
    \hline
    $f(x)$           &         1.&0e-04         &         3.&9e-06         &         6.&1e-07         &         2.&2e-08         \\ 
    $f^1(x)$         &         2.&4e-03         &         1.&3e-04         &         1.&4e-05         &         7.&3e-07         \\ 
    $f^2(x)$         &         6.&1e-02         &         1.&6e-03         &         4.&1e-04         &         1.&5e-05         \\ 
    $f^3(x)$         &           &              &         3.&1e-02         &         7.&2e-03         &         3.&5e-04         \\ 
    $f^4(x)$         &           &              &           &              &         8.&5e-02         &         5.&0e-03         \\ 
    $f^5(x)$         &           &              &           &              &           &              &         4.&0e-02         \\ 
    \hline
  \end{tabular}
  \end{center}
  \caption{
    The RMSs of the differences between the function values of $\sin(x)$ and those of the interpolating polynomial.
    The degree refers to the polynomial function used for the interpolation and
    the superscript at $f^n(x)$ to the $n^{th}$ derivative of $f(x)$.
    \label{tab:test-3}
    }
\end{table}

\noindent
As can be seen from looking at individual rows in table \ref{tab:test-3},
the RMSs becomes smaller for a higher degree, $N$, of the interpolating polynomial.
This is due to the larger number of points ($N+1$) used for the interpolation.
This dependence also applies to the derivatives,
i.e.\ the calculation of any derivative becomes more accurate with the degree of the interpolating polynomial.
In other words, the intrinsic features of Neville's algorithm are transferred to the derivatives using the above extension.
It is also interesting to compare the RMS of $f^n$ at some degree $N$ with those of $f^{n+2}$ at degree $N+2$.
Due to the nature of the original function,
the two test functions are then the same (apart from their sign) and,
as follows from the equations \ref{eq:neville-extended-1} and \ref{eq:neville-extended-2},
the effective number of points used for the calculation are then also the same.
For example,
the RMS of $f$ at degree 2 is 1.0e-04 and $f^2$ at degree 4 is 4.1e-04 and
the RMS of $f^2$ at degree 2 is 6.1e-02 and $f^4$ at degree 4 is 8.5e-02.
Indeed, these values agree reasonably well.
This shows that the accuracy of the calculation of the derivatives is consistent with that of the standard polynomial interpolation.

\section{Conclusions}

An extension of Neville's algorithm is presented which includes the derivatives of the interpolating polynomial.
This extension is based on the application of the chain rule for differentiating compositions of functions to
the recurrence relations that constitute Neville's algorithm.
The results are found to be consistent and numerically stable.

\bibliographystyle{plain}
\bibliography{Neville}
  
\end{document}